\documentclass[aps,pra,preprint,notitlepage,superscriptaddress,showpacs,10pt]{revtex4-1} 
\usepackage{amsmath}    
\usepackage{mathtools} 
\usepackage{amsfonts}
\usepackage{bm}
\usepackage{amssymb}
\usepackage{appendix}
\usepackage{graphicx}   
\usepackage{xcolor}      
\usepackage{subfigure}  
\usepackage[separate-uncertainty=true]{siunitx}
\usepackage{cancel} 

\usepackage{changes} 
\usepackage{soul}
\usepackage[normalem]{ulem}
\mathchardef\mhyphen="2D
\begin{document}

\title{Orbital Frontiers: Harnessing Higher Modes in Photonic Simulators}

\author{Jiho Noh}
\affiliation{Sandia National Laboratories, Albuquerque, NM, 87185 USA}
\affiliation{Center for Integrated Nanotechnologies, Sandia National Laboratories, Albuquerque, NM, 87185 USA}
\author{Julian Schulz}
\affiliation{Physics Department and Research Center OPTIMAS, RPTU University  Kaiserslautern-Landau, Kaiserslautern D-67663, Germany}
\author{Wladimir Benalcazar}
\affiliation{Department of Physics, Emory University, Atlanta, Georgia 30322, USA}
\author{Christina J{\" o}rg}
\affiliation{Physics Department and Research Center OPTIMAS, RPTU University  Kaiserslautern-Landau, Kaiserslautern D-67663, Germany}
\email[]{cjoerg@rptu.de}

\date{\today}

\begin{abstract}
Photonic platforms have emerged as versatile and powerful classical simulators of quantum dynamics, providing clean, controllable optical analogs of extended structured (i.e., crystalline) electronic systems. While most realizations to date have used only the fundamental mode in each site, recent advances in structured light -- particularly the use of higher-order spatial modes, including those with orbital angular momentum -- are enabling richer dynamics and new functionalities. These additional degrees of freedom facilitate the emulation of phenomena ranging from topological band structures and synthetic gauge fields to orbitronics. In this perspective, we discuss how exploiting the internal structure of higher-order modes is reshaping the scope and capabilities of photonic platforms for simulating quantum phenomena.
\end{abstract}

\maketitle


\section{Introduction \label{Introduction}}
Light has become an increasingly powerful medium for exploring interference effects and transport phenomena in extended structured systems (i.e., systems with extensive degrees of freedom), making photonic platforms valuable model systems for studying condensed matter physics. In solids, atoms form periodic lattices, and the electrons bound to them no longer behave as if they were isolated. Instead, their energy levels merge into bands, and the resulting band structure determines whether a material is a conductor, an insulator, or something more exotic, such as a superconductor or topological insulator. Studying these effects directly in real materials is extremely difficult, since they contain astronomically many atoms and parameters cannot easily be tuned. To overcome this, physicists turn to simplified lattice models -- mathematical descriptions (Hamiltonians) that capture the essential physics. Photonic systems provide an ideal platform to implement such lattice Hamiltonians that can be described by tight-binding models: by arranging “photonic atoms” such as waveguides or micropillars into lattices, one can mimic the behavior of electrons in crystals.

To build intuition, consider a one-dimensional system where a single photonic site -- such as a waveguide, micropillar, or cavity -- plays the role of a potential well. Each site hosts one or more modes with well-defined eigenvalues. Depending on the platform, these eigenvalues may correspond to a propagation constant (in waveguides) or a resonance frequency/energy (in cavities and polariton lattices). When two sites are brought close together, their modes hybridize, as light in these structures can couple between neighboring sites: the original states split into symmetric and antisymmetric superpositions, analogous to bonding and antibonding orbitals in quantum mechanics. Extending this principle to many sites arranged in a lattice gives rise to photonic bands, whose structure is set by the lattice geometry and coupling strengths, mirroring the formation of electronic bands in solids.
For example, take the case of a simple one-dimensional chain of five sites with one internal degree of freedom per site. In position space, the Hamiltonian can then be written as a $5 \times 5$ matrix, where the entries on the diagonal represent the on-site potentials, and the entries on the off-diagonals represent the coupling amplitudes between different sites.
This mapping is possible because the equations that govern light evolution in structured media are mathematically analogous to the Schrödinger equation. 
Thus, by carefully engineering spatial, temporal, and internal degrees of freedom, optical platforms can emulate the behavior of particles evolving under tailored Hamiltonians, providing insight into phenomena such as quantum transport~\cite{longhi_quantumoptical_2009,szameit_discrete_2010,grafe_integrated_2016}, localization~\cite{tal_schwartz_transport_2007, vaidya_reentrant_2023,stutzer_photonic_2018,cherpakova_transverse_2017,garanovich_light_2012}, topological phases~\cite{haldane2008possible, schulz_topological_2021,xie_photonics_2018,Iwamoto:21,Price_2022,Solnyshkov:21, Kremer:21,Lu_NatPhoton_2014,Ozawa_RevModPhys_2019, klembt_Nature_2018} and synthetic gauge fields~\cite{aidelsburger_artificial_2018,Joerg_LightSciAppl_2020,lumer_light_2019,cohen_generalized_2020,mittal_topologically_2014,lin_light_2014,umucalilar_artificial_2011}. These \emph{classical simulators of quantum dynamics} offer remarkable control, enabling the study of quantum-inspired dynamics in clean and tunable environments often inaccessible in traditional condensed matter systems.

Early implementations focused on the fundamental mode of light in each site, for example, in waveguide arrays to emulate single-particle dynamics in crystals. However, waveguides can host a rich internal mode structure. Beyond the fundamental mode, higher-order modes, including those carrying orbital angular momentum (OAM), provide additional degrees of freedom. These \emph{structured modes} resemble atomic orbitals in shape and symmetry (Fig.~\ref{fig1}) and can encode internal states, mediate inter-orbital couplings, and generate synthetic gauge fields.
In a parallel development, \emph{structured light} has offered dynamic, reconfigurable control. By shaping light with tools such as spatial light modulators, one can tune modal content and distribution in photonic lattices on demand. This provides full control over which sites and which modes are initially excited, enabling programmable orbital coupling and tailored state excitation.



In this perspective, we specifically focus on photonic simulators that combine two essential ingredients: a spatial structure of discrete sites, such as coupled waveguide arrays or polariton lattices, and the presence of multiple internal modes within each site.
Of course, multimode physics is not restricted to spatially structured simulators. Optical fibers, free-space cavities, and frequency-domain multimode platforms have already enabled remarkable advances in nonlinear optics, multimode lasing, quantum optics, communications, and photonic computing (see e.g. Ref.~\cite{Cristiani_2022,Wright:22}). There is a rapidly growing literature on multimode quantum optics and computational approaches that harness the large Hilbert spaces of multimode photonic systems. Yet, a comprehensive discussion of these directions lies outside the scope of this perspective.

Instead, we concentrate on two platforms that exemplify the potential of higher-order modes in each site of a lattice: photonic waveguide arrays and exciton-polariton microcavities.
In waveguide lattices, light propagation in the paraxial regime follows the paraxial wave equation that closely resembles the time-dependent Schr\"{o}dinger equation. This analogy allows for simulating quantum mechanical phenomena by studying the spatial propagation of light in classical optics. By carefully controlling the refractive index profile and coupling strengths between waveguides, various quantum phenomena have been studied, including Bloch oscillations~\cite{Pertsch_PhysRevLett_1999,stutzer_observation_2017,corrielli_fractional_2013,dreisow_bloch-zener_2009}, disordered systems and Anderson localization~\cite{tal_schwartz_transport_2007,stutzer_photonic_2018,garanovich_light_2012}, and topological effects~\cite{Ozawa_RevModPhys_2019}, within the realm of classical optics. Exciton-polariton microcavities, in contrast, rely on strong photon-exciton coupling to create hybrid quasiparticles called exciton-polaritons, which are governed by the driven-dissipative Gross-Pitaevskii equation. These systems consist of a quantum well placed between dielectric Bragg gratings, where excitons, which are bound electron-hole pairs, interact with confined photons to form polaritons. Polariton lattices allow to study many-body phenomena, such as Bose-Einstein condensation and superfluidity, as well as nonequilibrium topological phases such as polariton-based topological insulators.


Both platforms offer a level of control and observability rare in traditional condensed matter experiments: defects, geometry, coupling strength, and boundaries can be engineered on demand, and temporal evolution can be directly imaged through spatial propagation in waveguides or optical emission in cavities. This enables systematic exploration of single-particle effects in fundamental models and dynamic tuning of system parameters to emulate quantum dynamics in clean, versatile settings.

In this perspective, we examine how orbital degrees of freedom and structured light enrich photonic platforms. We begin with waveguide-based classical simulators of quantum dynamics, where higher-order spatial modes provide powerful routes to engineer coupling, symmetry, and synthetic fields beyond the ground-state paradigm.

\section{Orbitals in photonic waveguide lattices \label{waveguide arrays}}

\begin{figure}
    \centering
    \includegraphics[width=\linewidth]{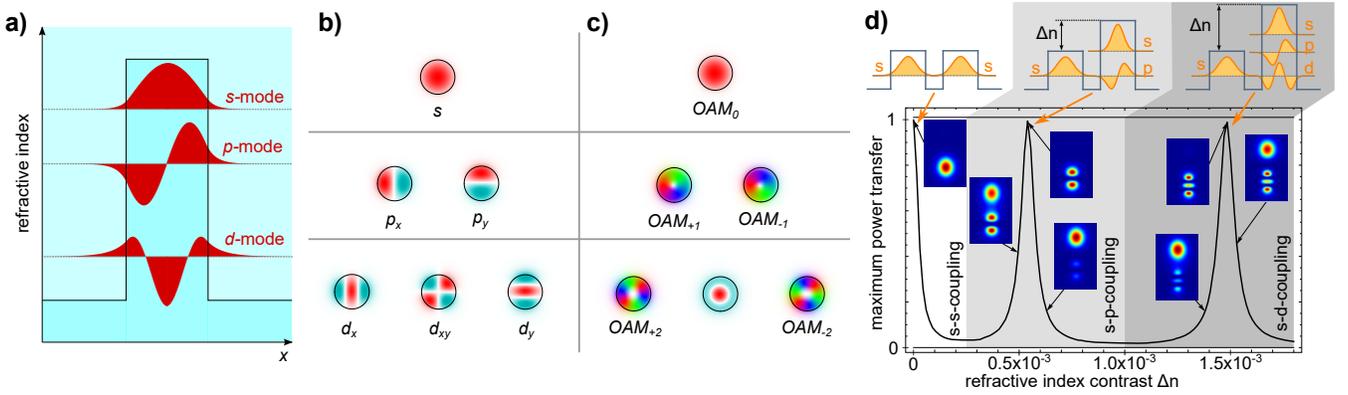}
    \caption{Orbital degrees of freedom in waveguide lattices.
    {a) In a quantum mechanical potential well, modes are classified according to their symmetry. The same can be done in step-index waveguides, where the effective refractive index plays the role of negative energy (i.e., the well is flipped). Due to this analogy, we can label} different Hermite-Gauss (HG) modes according to the notation used for atomic orbitals (b). Blue and red color stand for phases with a difference of $\pi$.
    c) Modes carrying a winding phase front, so-called OAM-modes, are composed of superpositions of HG modes with complex coefficients. 
    d) Coupling between different orbitals can be achieved when their propagation constants $\beta$ match. The propagation constants can be tuned via the refractive index difference $\Delta n$ between two waveguides, or the shape of the waveguides.
    d) Adapted with permission from \cite{Guzman-Silva}. Copyrighted by the American Physical Society.
    }
    \label{fig1}
\end{figure}

Photonic waveguide arrays provide an experimentally accessible setting to emulate quantum dynamics~\cite{longhi_quantumoptical_2009, szameit_discrete_2010}. 
Structurally, waveguide arrays consist of closely spaced dielectric waveguides, where light is confined to the high-index core but can tunnel into neighboring waveguides through evanescent coupling. Each waveguide acts as an artificial lattice site, while the coupling between adjacent waveguides mimics the quantum tunneling of particles separated by a potential barrier. The fundamental analogy underpinning this approach lies in the mathematical correspondence between the paraxial Helmholtz equation for light propagation in a weakly guiding medium 
\begin{equation*}
    i \partial_{\textcolor{red}{z}} E_\mathrm{el}(x,y,\textcolor{red}{z})=\left[-\frac{\nabla_\perp^2}{2 n_0 k_0} -{k_0}\textcolor{blue}{\Delta n(x,y,z)}\right] E_\mathrm{el}(x,y,\textcolor{red}{z}),
\end{equation*}
and the time-dependent Schr\"{o}dinger equation
\begin{equation*}
    i \partial_{\textcolor{red}{t}} \Psi(x,y,\textcolor{red}{t})=\left[-\frac{\nabla_\perp^2}{2m} +\textcolor{blue}{V(x,y,t)}\right] \Psi(x,y,\textcolor{red}{t})
\end{equation*}
\cite{szameit_discrete_2010, longhi_quantumoptical_2009}, with the bulk refractive index $n_0$, refractive index variation $\Delta n(x,y,z)=n_0-n(x,y,z)$, wave vector in vacuum $k_0$, mass $m$, and we have set $\hbar=1$. 
In this analogy, the longitudinal propagation direction $z$ of the light takes the role of time, while variations in the transverse refractive index $n(x,y)$ take the role of an effective potential $V$ in the analogous Schr\"{o}dinger-like equation. 

This optical analogy enables the direct simulation of dynamics over a range of physical systems. For example, waveguide arrays have been used to model adiabatic state transfer (e.g., STIRAP)~\cite{longhi_quantumoptical_2009}, explore Dirac cones and edge states in graphene-like lattices~\cite{plotnik_observation_2014,rechtsman_topological_2013,song_unveiling_2015,Noh_PhysRevLett_2018}, and realize topological insulators~\cite{rechtsman_FTI_2013, maczewsky_observation_2017,jorg_dynamic_2017,Noh_NatPhoton_2018}. These experiments are typically implemented using single-mode waveguides, where each waveguide supports only the fundamental mode, providing a straightforward mapping to tight-binding models with a single electron per atomic site. However, real condensed matter systems are far more intricate: atoms may host multiple higher-energy orbitals (such as $p$- and $d$-orbitals), whose angular structures allow electrons to carry orbital angular momentum, which is a key ingredient of many phenomena, starting with how an electron bonds with other electrons to form a molecule.

Therefore, recent work in photonic simulators has begun to explore the rich, underutilized modal structure available in multimode waveguides (see also~\cite{vicencio_multi-orbital_2025}). These higher-order spatial modes can be used to mimic atomic orbitals or synthetic dimensions. 

Individual dielectric waveguides can support multiple transverse eigenmodes. The number of supported modes in a cylindrical step-index waveguide is determined by its V-number -- a function of core radius, wavelength, and refractive index contrast. In analogy to quantum mechanical potential wells, we can describe these modes in terms of their symmetry, where the mode with the highest propagation constant $\beta$ (equivalent to highest effective refractive index) has a Gaussian-like intensity profile. Modes subsequently lower in $\beta$ have an increasing number of nodes, i.e., they exhibit $\pi$-changes in their phase-profile (Fig.~\ref{fig1}~a). Within the framework of Hamiltonian simulation, these modes are frequently labeled using atomic orbital terminology: the fundamental Hermite-Gauss mode $\mathrm{HG}_{00}$ is referred to as the $s$-mode, while the first-order higher modes with nodal planes are called $p_x$ and $p_y$, corresponding to $\mathrm{HG}_{10}$ and $\mathrm{HG}_{01}$, respectively (Fig.~\ref{fig1}~b). Note that the polarization of the electric and magnetic field, which can be mapped to a spin degree of freedom, is often neglected due to the very low spin-orbit coupling in photonic systems with low refractive index contrast.
In radially symmetric waveguides, an alternative basis to the HG modes can be formed by modes carrying orbital angular momentum (OAM). These are modes with an azimuthal phase term $\exp{(i l\phi)}$ exhibiting a winding phase front (Fig.~\ref{fig1}~c). Here, $l$ is the OAM quantum number. By basis transformation, OAM modes can be composed of superpositions of HG modes with complex coefficients: for example, an OAM mode with $l=\pm 1$ is given by $p_x \pm i p_y$.

\subsection{Inter-orbital coupling \label{Inter-orbital coupling}}
Orbital modes form an orthogonal basis within a single waveguide, and, due to their differing propagation constants $\beta$, typically do not couple between neighboring waveguides. However, just as the energy levels in a quantum potential well can be tuned by altering its shape, the modal propagation constants in a waveguide can be engineered through modifications to the geometry and refractive index contrast of the structure. For instance, increasing $\Delta n$ (the index contrast between core and cladding) raises the confinement of higher modes and can bring the $\beta$-value of a $p$-mode closer to that of an $s$-mode in a neighboring, unchanged, waveguide. This tuning enables inter-orbital coupling across waveguides~\cite{Guzman-Silva} (Fig.~\ref{fig1}~d).
Notably, this coupling is not limited to static refractive index engineering. In Kerr-nonlinear waveguides, intensity-dependent changes in the refractive index can dynamically modify modal propagation constants. Due to the distinct spatial field distributions of $s$- and $p$-modes, they experience different degrees of nonlinear index shift~\cite{rajeevan_nonlinear_2025}. Nonlinear interactions can thus be harnessed to enable intensity-tunable coupling between orbitals. This is particularly exciting because it allows the system’s effective Hamiltonian to be reconfigured in real time, opening the door to exploring nonlinear and mode-dependent phenomena that are impossible in static structures. 

Beyond refractive index tuning, the shape of the waveguide cross-section, such as its ellipticity, plays a crucial role in defining the relative energy levels of supported modes. By elongating the core along a particular axis, the propagation constants of the corresponding eigenmodes can be shifted.
Furthermore, supersymmetry transformations in photonic systems have also been applied to engineer mode-conversion between paired optical modes~\cite{Viedma:21}. The novelty lies in designing synthetic structures where one mode can be transformed into its supersymmetric partner, enabled by refractive-index landscapes that map their propagation constants, allowing controlled reshaping or routing of modes within photonic lattices.

\subsection{Orbitals can induce phases in the coupling \label{phases in coupling}}
\begin{figure}
    \centering
    \includegraphics[width=\linewidth]{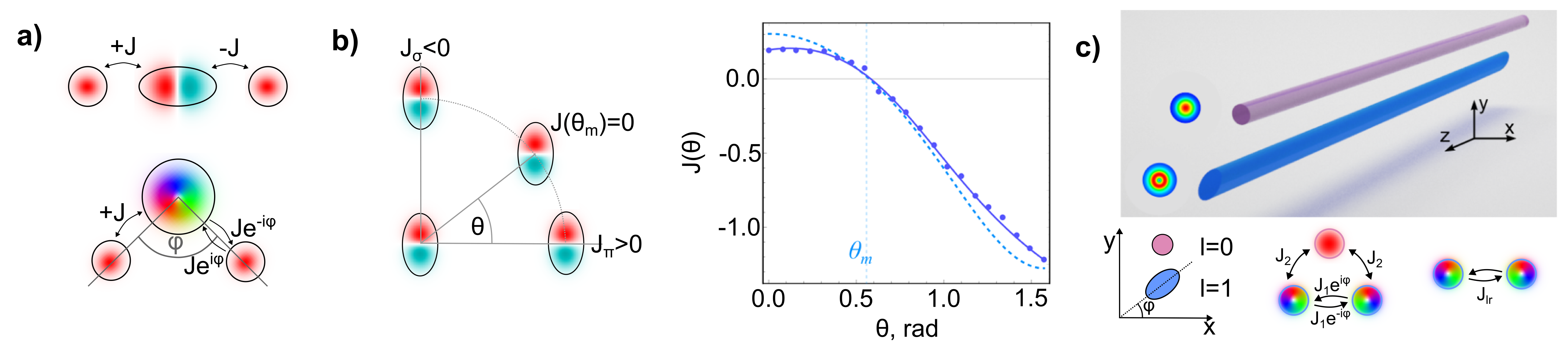}
    \caption{Phases in the coupling amplitude appear depending on the orientation of the orbitals in the transverse plane.
    a) The coupling amplitude $J$ of an $s$- to a $p_y$- and back to an $s$-mode gains a $\pi$-phase flip, i.e. a negative sign (top). The coupling of $s$- and OAM-modes depends on the angle between waveguides and the OAM quantum number (bottom).
    b) The phase in the coupling between two $p_y$-modes depends on the angle between them. It can range from positive coupling, when two $p_y$-modes are aligned along the x-axis (coupling $J_\pi$), to a negative one for alignment along the y-axis (coupling $J_\sigma$). At a certain ``magic" angle $\theta_m$, the coupling vanishes completely.
    c) Coupling between a detuned and lossy $s$-mode waveguide and an elliptical $p$-mode waveguide has been predicted to lead to non-Hermitian states~\cite{wang_non-hermitian_2023}.
    b) Adapted with permission from \cite{invisibility}. Copyright 2025 American Chemical Society. c) Adapted with permission from~\cite{wang_non-hermitian_2023} © Optica Publishing Group. 
    }
    \label{fig2}
\end{figure}
One particularly intriguing consequence of inter-orbital coupling is the induced phase in the coupling, which depends on the direction along which waveguides are arranged. For instance, coupling an $s$-mode in one waveguide to a $p_x$-mode in another and back to an $s$-mode in a third, where the three waveguides are arranged along the $x$-axis, introduces a relative $\pi$-phase shift (i.e. a negative sign) in the coupling amplitude (Fig.~\ref{fig2}~a).

More broadly, the coupling phase between orbitals is determined by the relative angle between waveguides. For example, the coupling between two $p_y$-modes is positive when the orbitals are aligned along the $x$-axis but becomes negative when aligned along the $y$-axis. The phase is thus determined by the relative orientation of the waveguides and can be tuned (Fig.~\ref{fig2}~b) \cite{invisibility}. In particular, there are specific relative orientations where the coupling amplitude vanishes altogether. This property has been harnessed to design flat bands and suppress nearest-neighbor hopping in deformed photonic graphene lattices~\cite{invisibility}.

Even richer phenomena emerge when losses are introduced into the system. In recent simulations, coupling between OAM modes has been shown to exhibit non-Hermitian behavior under carefully engineered configurations \cite{wang_non-hermitian_2023}. The studied system consists of a single elliptical multimode waveguide (with long axis $\rho r$ and short axis $r/\rho$) coupled to a lossy single-mode waveguide (Fig.~\ref{fig2}~c). In the regime of large detuning $\delta$ between their propagation constants, the single-mode waveguide can be adiabatically eliminated, yielding an effective non-Hermitian Hamiltonian. For a lossy single-mode waveguide with loss $\gamma$, such that $\delta = i\gamma$, and for $\rho=1$, the system realizes an anti-PT symmetric configuration, which suppresses the symmetric superposition of OAM modes. A change to $\rho = 1.02$ and an angular orientation $\varphi = \pi/4$ instead results in a PT-symmetry breaking transition, highlighting the sensitivity of modal coupling to both geometry and loss.

\subsection{Synthetic gauge fields \label{AGF}}
\begin{figure}
    \centering
    \includegraphics[width=\linewidth]{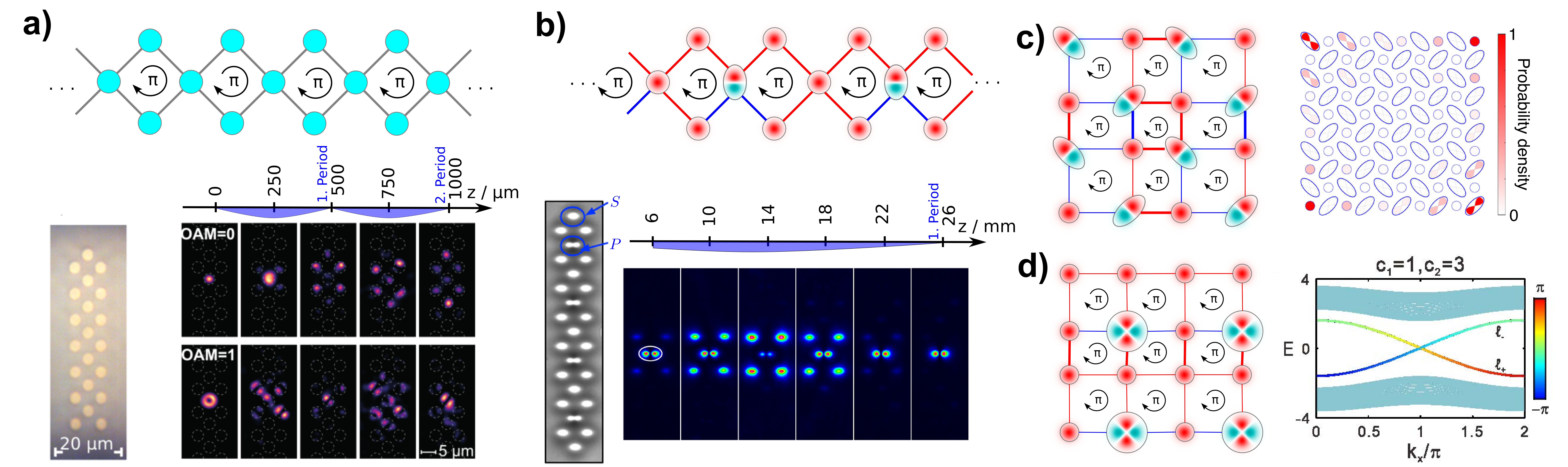}
    \caption{Gauge fields induced via higher orbitals or OAM in photonic waveguide lattices.
    a) In a diamond chain, light with OAM of $|l|=1$ creates a $\pi$-flux, leading to Aharonov-Bohm caging (a return of the intensity to the excited waveguide after a certain propagation distance period), while light with vanishing OAM disperses~\cite{Joerg_LightSciAppl_2020}. 
    b) Similarly, $p_y$-modes in a diamond chain create a flux of $\pi$~\cite{PhysRevLett.128.256602}.
    c) Elliptical waveguides in a 2D SSH lattice create fluxes that lead to the formation of corner states in higher-order topological insulators~\cite{Schulz_NatCommun_2022}.
    d) $d$-modes in a 2D lattice give rise to a topological M\"{o}bius insulator~\cite{Jiang_OpticsLetters_23}
    b) Adapted with permission from~\cite{PhysRevLett.128.256602}. Copyright 2022 by the American Physical Society. d) Adapted with permission from ~\cite{Jiang_OpticsLetters_23}. © Optica Publishing Group. 
    }
    \label{fig3}
\end{figure}

Adding complex phase factors to the coupling between different orbitals presents a powerful method for generating synthetic flux within photonic waveguide arrays. 
Flux is present when the phases accumulated in the hopping along a closed path in the lattice do not amount to integer multiples of 0 or $2\pi$. Therefore, flux can be achieved by carefully controlling the hybridization of different orbital modes, such as $s$- and $p$-orbitals or $s$- and $d$-orbitals.  This method enables the realization of photonic bands in various topological phases.
Schulz et al.~\cite{Schulz_NatCommun_2022} demonstrated experimentally a photonic quadrupole topological insulator by using $s$- and $p$-orbital modes. These modes were implemented as eigenmodes in waveguides fabricated using direct laser writing in circular (for $s$-modes) and elliptical (for $p$-orbitals) shapes to induce synthetic $\pi$ magnetic flux in a plaquette (Fig.~\ref{fig3}~c). The non-trivial topology via this approach was confirmed by observing topologically-protected zero-dimensional corner states~\cite{benalcazar2017quantized}.
Extending this multiorbital approach, researchers have theoretically engineered photonic M\"{o}bius topological insulators in waveguide arrays using inter-orbital coupling to generate synthetic gauge flux.
By coupling different orbital modes -- $s$ and $d$ (Fig.~\ref{fig3}~d)~\cite{Jiang_OpticsLetters_23} or $s$ and $p$~\cite{Liu_Nanophotonics_2023} -- they created a synthetic $\pi$ flux, leading to a projective translation symmetry. Breaking this symmetry transforms the system into a topological insulator with M\"{o}bius twisted edge bands. The 4$\pi$ periodicity of these bands leads to two twisted edge states for each transverse Bloch momentum which can be selectively stimulated for opposite beam transport. Additionally, Z. Liu et al.~\cite{Liu_Nanophotonics_2023} showed that this technique can transform a Dirac semimetal into a Weyl-like semimetal with flat bands.

In rotationally symmetric waveguides, a natural modal basis consists of OAM modes, which possess azimuthally varying phase fronts and ring‑shaped intensity, carrying an integer topological charge $l$, encoding the quantized angular momentum of the mode. This spatial phase structure makes OAM an effective tool for inducing synthetic gauge fields in photonic systems and mimicking the behavior of charged particles in magnetic fields.
%
The coupling between two OAM modes of differing topological charge $l_1$ and $l_2$ introduces a phase factor that scales as $\propto \exp(i(l_1–l_2)\varphi)$ with the azimuthal angle $\varphi$ between waveguides (Fig.~\ref{fig2}~a) \cite{turpin_engineering_2017,Joerg_LightSciAppl_2020}. In certain lattice geometries, such as a rhombic lattice (also known as a diamond chain), this phase structure enables the realization of synthetic magnetic fields. 

In this context, J\"{o}rg et al.~\cite{Joerg_LightSciAppl_2020} experimentally demonstrated synthetic gauge fields by injecting first-order OAM modes ($l = \pm1$) into a diamond-chain lattice of direct laser-written optical waveguides. They showed that the system can effectively be decoupled into two sets with $\pi$-fluxes threading the plaquettes of each \cite{Joerg_LightSciAppl_2020}. The prevalence of this $\pi$-flux was confirmed through Aharonov-Bohm (AB) caging experiments (Fig.~\ref{fig3}~a), where interference due to the flux leads to a periodic return of the intensity to the central waveguide (so-called caging). In contrast, injected light without OAM disperses transversely. This highlights that higher-order and OAM modes allow one to dynamically and externally tune the properties of a given photonic lattice, just by changing the light input. This dynamic control paves the way for scalable and reconfigurable topological photonic devices~\cite{Joerg_LightSciAppl_2020, Wang_PhysRevA_2024}. The same caging effect can also be achieved by replacing every second central site with a $p$-mode waveguide instead of using OAM modes (Fig.~\ref{fig3}~b) \cite{PhysRevLett.128.256602}.

Following this work, Wang et al.~\cite{Wang_PhysRevA_2024} and Jiang et al.~\cite{Jiang_JLightTechnol_2023} theoretically explored complementary methods. Wang et al. proposed an alternative method to generate synthetic gauge fields using the hybridization of fundamental ($l = 0$) and first-order OAM modes in a zigzag array. Jiang et al. also theoretically investigated topological bound modes with first-order OAM modes in zigzag waveguide arrays, effectively mimicking the Creutz ladder model~\cite{Jiang_JLightTechnol_2023}. Their research highlighted that topological propagation can be uniquely influenced by OAM, and they also presented the orientation angle of elliptical waveguides as an extra degree of freedom, providing an enhanced control method for these topological bound modes.
OAM modes on a zigzag chain with tunable angle between waveguides have also been suggested to realize arbitrary fluxes on a photonic diamond chain lattice, which can be decorated with impurities. This leads to a flux-mediated Su–Schrieffer–Heeger (SSH) model, demonstrating how tuning impurity placement and flux enables engineering topologically nontrivial band structures and robust edge modes~\cite{viedma_flux-mediated_2024}.

In addition to synthetic fluxes, the inherent anisotropy of $p$-orbitals alone has allowed for the realization of a variety of topological phenomena, first explored in polariton lattices.
While the anisotropy of $p$-orbital hopping has been extensively studied in polariton systems, its potential extensions in waveguide-based platforms remain largely unexplored.
Recent experiments have demonstrated the utility of $p$-orbital modes in realizing higher-order topological bands in laser-written optical waveguide arrays~\cite{Zhang_eLight_2023, Bongiovanni_LaserPhotonReview_2024}.

\section{Higher modes in polariton lattices}
The orbital degree of freedom provides an additional dimension in platforms where photon-photon interactions are non-negligible. 
For example, higher modes also play a significant role in lattices of exciton-polariton condensates, where the interaction of the excitonic component of the polariton affects the photonic component.

\begin{figure}
    \centering
    \includegraphics[width=\linewidth]{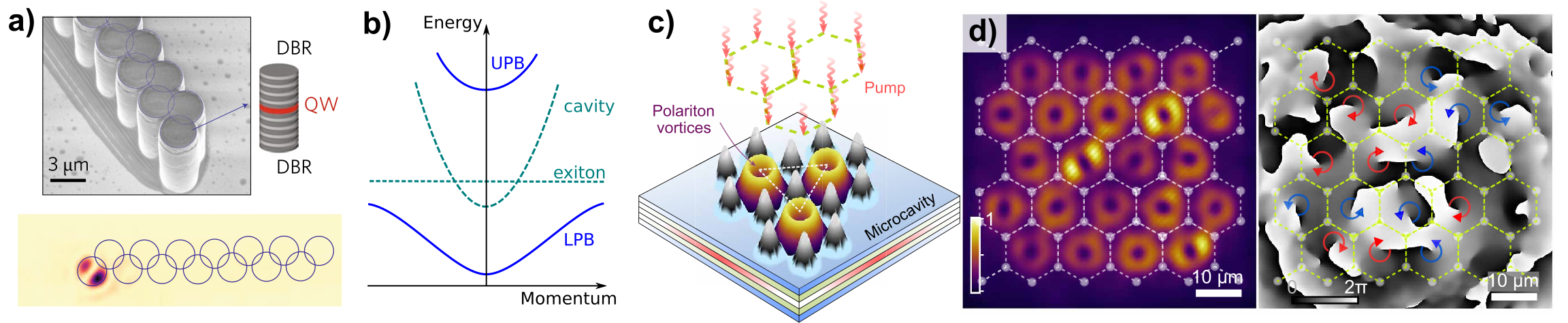}
    \caption{Higher modes in polariton lattices.
    \textbf{a)} A zig-zag array of coupled polariton cavities with $p$-modes acts as a Su-Schrieffer-Heeger system. The sketch on the side shows the different layers of an exciton-polariton cavity, a quantum well (QW) in an optical cavity consisting of two dielectric Bragg gratings (DBR). The picture on the bottom shows the light emission of the spectrally isolated and localized topological edge mode~\cite{St-Jean_NatPhoton_2017}.
    \textbf{b)} Dispersion relation: The dispersion relation of a photon in the cavity is parabolic, while that of an exciton is, in comparison, flat. Due to the coupling between the two particles, the polariton dispersion relation splits into an upper and a lower polariton branch (UPD and LPB, respectively).
    \textbf{c)} Sketch of how a structured pump beam acts as a repulsive potential for polaritons. Lattices can be constructed such that polaritons populate vortex modes on each lattice site. \cite{Alyatkin_SienceAdvances_2024}.
    \textbf{d)} Measured emission (intensity right and phase left) of a polariton vortex lattice. The OAM of the vortex modes arrange themselves in the same way as spins would do in an Ising lattice.~\cite{Alyatkin_SienceAdvances_2024}.
    a) Reprinted with permission from~\cite{St-Jean_NatPhoton_2017}, Copyright © 2017 Springer Nature.
    c), d) Copyright © 2024, The American Association for the Advancement of Science. 
    }
    \label{fig:polariton}
\end{figure}

An exciton-polariton is a quasiparticle that behaves partly like a photon and partly like an electron-hole pair.
To investigate them, a microcavity consisting of a quantum well (QW) surrounded by dielectric Bragg gratings (DBG) on both sides, acting as cavity mirrors, is commonly used (Fig.~\ref{fig:polariton}~a).
The space between the cavity mirrors is so small that only a single longitudinal light mode can be excited in the quantum well.
Therefore, the longitudinal mode $k_z$ of the photons inside the cavity is fixed, and the dispersion relation of the photon is that of a (very light) massive particle.
The rest mass of the photon -- the bottom of its dispersion relation -- can be tuned via varying the cavity length.

Inside these semiconductor quantum well, excitons can be excited.
An exciton is an electron-hole pair in an insulator or a semimetal that is bound together by the Coulomb force.
Due to the presence of a bandgap, excitons emit photons when they decay.
Since the mass of the exciton is much larger than the effective mass of the photon, its dispersion relation is much flatter (Fig.~\ref{fig:polariton}~b). 
If the photon energy and the exciton energy are close, strong light-matter interaction occurs, and polaritons -- the superposition of photons and excitons -- are formed.
Here, the polariton dispersion relation splits into an upper and a lower branch due to the strong coupling.
Because of the excitonic fraction of the polariton, polaritons can interact with each other via the Coulomb force and exchange momentum and energy, and thermalize into a ground state~\cite{Kavokin_NatPhysRev_2022}.
Also, due to the excitonic fraction, polaritons react to a magnetic field, as shown in~\cite{klembt_Nature_2018}.
Polaritons in a cavity can be excited either optically or electrically.

In platforms supporting polaritons, lattice structures can be implemented by structuring the cavity (often done by etching parts of the upper mirror as seen in Fig.~\ref{fig:polariton}~a) or by pumping the cavity with a structured pump beam. 
When the cavity is structured, the diameter of the individual lattice points can be chosen large enough so that higher modes, such as $p$-modes and $d$-modes, can be excited~\cite{St-Jean_NatPhoton_2017,Klembt_APL_2017,Scafirimuto_CommunPhys_2021}.
In particular, the transformation from isotropic $s$-orbital modes to anisotropic $p$-orbital modes further demonstrates the versatility of polariton systems, enabling richer and more complex band structures even without synthetic gauge fields. 
Key discoveries include the emergence of Dirac cones and a flatband in a honeycomb lattice~\cite{Jacqmin_PhysRevLett_2014}, the demonstration of $p$-orbital Su–Schrieffer–Heeger (SSH) edge states and their lasing behavior~\cite{St-Jean_NatPhoton_2017, Harder_ACSPhotonics_2021} (see Fig.~\ref{fig:polariton}~a), and the realization of $p$-orbital edge states in a 2D photonic honeycomb lattice~\cite{Milicevic_PhysRevLett_2017}. 
Significantly, photonic $p$-orbital graphene was the platform for the first experimental observation of critically tilted (Type-III) Dirac cones, in addition to semi-Dirac and tilted Type-I Dirac cones~\cite{Milicevic_PhysRevX_2019}. 

In contrast to etched structures, if the potential landscape is created by the structure of the pump beam, the potential can be adjusted dynamically. 
Tosi et al. showed that the pump beam can be used, not solely as a particle source, but also in order to define a potential~\cite{Tosi_NatPhys_2012}.
This is possible due to the repulsive interaction between polaritons that can be mathematically described in a mean-field approximation as a repulsive density-dependent potential.
{Where the cavity is pumped locally, a high polariton density is created. This high polariton density acts as a repulsive potential, which can cause the polaritons to settle in the spaces between the pumped areas during thermalization
~\cite{Tosi_NatPhys_2012,Askitopoulos_PhysRevB_2013}.
Alyatkin et al. showed later that it depends on the pump spot geometry whether polaritons condense at the pump spots or in between, and created a lattice structure that hosted $p$-orbital flat bands~\cite{Alyatkin_NatCommun_2021}.
By selectively pumping cavities and creating these repulsive potentials, polaritons in stable vortex modes with OAM $=\pm1$ and/or circular polarization were experimentally excited and manipulated~\cite{Ma_NatCommun_2020,Ohadi_PhysRevX_2015}. 
The modes with OAM $=\pm1$ are stable since they represent an attractor in these structured polariton systems. The generation and precise manipulation of stable vortices is important with respect to using polariton vortices in information processing.

The excitation of modes with higher OAM has also been investigated theoretically~\cite{Barkhausen_OpticsLetters_2020,Pukrop_PhysRevB_2020}.
However, even with systems that aim for OAM $\pm1$ modes, it is possible to investigate optically complex systems that cannot be replicated with $s$-mode lattice systems.
For example, it was demonstrated that in lattices of coupled polariton vortex modes (Fig.~\ref{fig:polariton}~c and d), the OAM modes orient themselves in the same way as spins arrange in an Ising lattice. 
The authors also showed that by varying the coupling between the vortex sites, they can choose to observe a ferromagnetic or antiferromagnetic arrangement~\cite{Alyatkin_SienceAdvances_2024}.


\section{Perspective \label{Outlook}}
Having reviewed current implementations of higher modes in photonic simulators of quantum dynamics, we now present our perspective on the emerging directions and future opportunities in the field.

Looking ahead, orbital modes in photonic platforms with many sites offer a promising route to explore concepts from orbitronics -- the study and control of orbital degrees of freedom for information processing in solid-state systems. In electronic materials, orbital transport is often intertwined with spin and charge dynamics, making it difficult to isolate and study the pure role of orbital motion. In contrast, photonic waveguide lattices naturally support structured light modes that resemble atomic orbitals, but are free from spin-orbit coupling and Coulomb interactions. This makes them ideal platforms to disentangle orbital-specific effects from other degrees of freedom.
Higher orbitals can enable versatile device functionalities with topological protection, especially when combined with synthetic dimensions~\cite{Lustig_AdvOptPhoton_2021} and nonlinear effects~\cite{Smirnova_ApplPhysRev_2020}. Theoretical frameworks based on symmetry indicators and band representations~\cite{Bradlyn_Nature_2017,po2017} have already categorized phases that are intimately related to orbital degrees of freedom, leading to the realization of multipolar and higher-order topological bands~\cite{Mazanov_PhysRevB_2024,Liu_PhysRevB_2024}.

By selectively exciting and coupling orbital modes, one can emulate orbital currents, design lattices with orbital-dependent hopping, and possibly realize analogs of orbital Chern insulators -- topological phases driven solely by orbital motion. Recent theoretical work has also begun to explore the interface between non-Hermitian and orbital systems. This suggests that structured photonic lattices could also be used to probe exotic effects like the orbital skin effect, where breaking time reversal and parity symmetries could lead to orbital-dependent nonreciprocal dynamics and boundary localization.

Orbital modes may also shape defect-bound states. Topological defects carrying torsion or curvature singularities can confine modes at their core, typically stabilized by a bulk gap~\cite{teo2010}. Controlling the orbital character of these modes could enforce symmetry mismatch with bulk states, enabling confinement even without a gap. This suggests a new class of defect-bound states in the continuum~\cite{benalcazar2020bound,vaidya2021point}. Breaking symmetries could then turn them into tunable resonances, providing a mechanism for controllable coupling.

Kerr nonlinearity provides another powerful route. Recent advances in multimode nonlinear photonics may expand the degree of control over orbital modes. Reconfigurable platforms and nanofabricated structures could allow dynamic tuning of orbital coupling and synthetic gauge fields. Additionally, computational tools, such as physics-informed machine learning and optimization algorithms developed for multimode systems, can be adapted to design orbital-specific couplings that lead to topological bands with high precision. 
Recent work also links band topology with solitons in the Gross--Pitaevskii equation, showing that solitons can inherit the position and symmetry representation of the Wannier orbital from which they bifurcate~\cite{jurgensen2023quantized,schindler2025nonlinear}. This enables orbital-selective excitation of soliton modes, offering new control of nonlinear dynamics.
At the same time, weak Kerr nonlinearities in fibers reveal conserved thermodynamic properties of mode populations~\cite{MarquesMuniz_Science_2023,Ren_PhysRevLett_2023,Podivilov_PhysRevLett_2022,Wu_PhysRevLett_2022}. Extending these ideas to coupled arrays could uncover collective effects such as mode-resolved thermalization, energy transport, and non-equilibrium steady states, providing insights into many-body dynamics.
These advancements may broaden the scope of orbital-mode research, leading to adaptable systems capable of exploring intricate quantum phenomena.

Dynamic modulation offers yet another direction. Time-periodic variations, such as shaping a waveguide along the propagation axis, can couple onsite orbital modes that would otherwise remain decoupled. This Floquet engineering allows effective Hamiltonians with tailored band structures, symmetry breaking, or synthetic gauge fields. Dynamically driven orbital hybridization could simulate exotic topological phases.
\\
\\
In conclusion, orbital degrees of freedom in photonic simulators are only beginning to reveal their potential. They open a rich landscape for devices with new functionalities and to emulate quantum dynamics. From synthetic gauge fields to orbital thermodynamics, these systems provide new opportunities for Hamiltonian simulation -- and much remains to be discovered.

\section*{Acknowledgments}
CJ and JS gratefully acknowledge financial support from the DFG through SFB TR 185 OSCAR, Project No. 277625399. WAB is thankful for the support of startup funds from Emory University. 
JN was supported by the U.S. Department of Energy (DOE), Office of Basic Energy Sciences, Division of Materials Sciences and Engineering. This work was performed, in part, at the Center for Integrated Nanotechnologies, an Office of Science User Facility operated for the U.S. DOE Office of Science. Sandia National Laboratories is a multimission laboratory managed and operated by National Technology and Engineering Solutions of Sandia, LLC., a wholly owned subsidiary of Honeywell International, Inc., for the U.S. DOE’s National Nuclear Security Administration under contract DE-NA0003525. This paper describes objective technical results and analysis. Any subjective views or opinions that might be expressed in the paper do not necessarily represent the views of the U.S. DOE or the United States Government.
ChatGPT was used for language editing purposes.

\bibliography{references}

\end{document}